\documentclass[aps,prb,reprint,superscriptaddress]{revtex4-1}
\usepackage[utf8]{inputenc}
\usepackage{amsfonts}
\usepackage{graphicx}
\graphicspath{{Pictures/}}
\usepackage{bm} % bold math
\usepackage{amssymb} % use this package to enable \nrightarrow command
\usepackage{amsmath} % use this package to enable \xrightarrow command
\usepackage{braket} % use for Dirac bra-kets : \rangle \labgle & \mid
\usepackage{natbib} % bibtex package
\usepackage[colorlinks = true, linkcolor = magenta, urlcolor  = purple, citecolor = purple, anchorcolor = blue]{hyperref}
\usepackage[T1]{fontenc}
\usepackage{xcolor}
\usepackage{marginnote}
\usepackage{array,multirow}
\usepackage{mathtools}
\usepackage{tabularx,booktabs}
\usepackage{MnSymbol}

\newcolumntype{P}[1]{>{\centering\arraybackslash}p{#1}}

\newcolumntype{Y}{>{\centering\arraybackslash}X}
\newcommand{\vk}{\ensuremath{\mathbf{k}}}

\newcommand{\floor}[1]{\lfloor #1 \rfloor}

\begin{document}

\title{Higher Chern numbers in Multilayer ($\mathcal{N} \ge 2$) Lieb Lattices: Topological Transitions and Quadratic Band Crossing Lines}

\author{Saikat Banerjee}
\email{saikat.banerjee@physik.uni-augsburg.de}
\affiliation{Theoretical Physics III, Center for Electronic Correlations and Magnetism, Institute of Physics, University of Augsburg, 86135 Augsburg, Germany,}
\author{Avadh Saxena}
\email{avadh@lanl.gov}
\affiliation{Theoretical Division and Center for Nonlinear Studies, Los Alamos National Laboratory, Los Alamos, New Mexico 87545, USA}

\date{\today}
%%%%%%%%%%%%%%%%%%%%%%%%%%%%%%%%%%%%%%%%%%%%%%%%%%%%%%%%%%%%%%%%%%%%%%%%%%%%%%%%%%%%%%%%%%%%%%%%%%%%%%%%%%%%%%%%%%%
%Abstract
%%%%%%%%%%%%%%%%%%%%%%%%%%%%%%%%%%%%%%%%%%%%%%%%%%%%%%%%%%%%%%%%%%%%%%%%%%%%%%%%%%%%%%%%%%%%%%%%%%%%%%%%%%%%%%%%%%%
\begin{abstract}
We consider a hitherto unexplored setting of stacked multilayer ($\mathcal{N}$) Lieb lattice which undergoes an unusual topological transition in the presence of intra-layer spin-orbit coupling (SOC). The specific stacking configuration induces an effective non-symmorphic 2D lattice structure, even though the constituent monolayer Lieb lattice is characterized by a symmorphic space group. This emergent non-symmorphicity leads to multiple doubly-degenerate bands extending over the edge of the Brillouin zone (i.e. Quadratic Band Crossing Lines). In the presence of intra-layer SOC, these doubly-degenerate bands typically form three $\mathcal{N}$-band subspaces, mutually separated by two band gaps. We analyze the topological properties of these multi-band subspaces, using specially devised Wilson loop operators to compute non-abelian Berry phases, in order to show that they carry a higher Chern number $\mathcal{N}$.
\end{abstract}
%%%%%%%%%%%%%%%%%%%%%%%%%%%%%%%%%%%%%%%%%%%%%%%%%%%%%%%%%%%%%%%%%%%%%%%%%%%%%%%%%%%%%%%%%%%%%%%%%%%%%%%%%%%%%%%%%%

\maketitle
%================================================================================================================================
\section{Introduction \label{sec:intro}}
%================================================================================================================================

The discovery of topological materials~\cite{Bernevig1757,PhysRevLett.96.106802} and their subsequent ten-fold symmetry classification within the Altland-Zirnbauer scheme~\cite{RevModPhys.82.3045,RevModPhys.83.1057}, has led to a rapid surge in the search for new materials with non-trivial topological properties. This primarily includes, for instance, (i) the time-reversal invariant topological insulators (TIs), commonly known as quantum spin Hall (QSH) insulators in two dimensions, which were predicted to occur in strongly spin-orbit coupled materials~\cite{Moore2010,RevModPhys.83.1057,PhysRevLett.95.226801}, (ii) the particle-hole symmetric band structures, which lead to interesting topological phases such as topological superconductors~\cite{PhysRevB.78.195125,kitaev.09,PhysRevLett.102.187001}, and (iii) more recently, the topological crystalline insulators (TCI), which are protected  by a combination of the time-reversal and the underlying point group symmetry of the associated lattice~\cite{PhysRevLett.106.106802,yoichi.15,PhysRevX.8.031070,PhysRevX.7.041069,Slager2013}. As these systems are generally understood within a single-particle picture, one typically characterizes their topology by associating a topological invariant (Chern number) to the resulting band structure. 

 Finding new materials with tunable Chern numbers ($\mathcal{C}$) is enormously important as they are directly measurable in terms of the quantized Hall conductance ($\mathcal{C}e^2/h$) of two-dimensional (2D) Chern insulators~\cite{PhysRevX.1.021014,PhysRevLett.112.046801}. In the presence of quasiparticle interaction, such integer quantization of conductance further breaks down into fractional values. In the case of relatively high Chern numbers, even potentially new phases (\textit{viz.} topological nematic phases~\cite{PhysRevX.2.031013}) can emerge due to the interplay of topology and strong quasiparticle correlations. Evidence of such high Chern numbers for almost flat-band systems has been discussed in previous theoretical works~\cite{Rachel_2018} on various lattices, including Kagome~\cite{PhysRevB.86.241111}, triangle~\cite{PhysRevLett.108.126405}, and checkerboard~\cite{PhysRevLett.106.236804}. More interestingly, it was shown in Ref.~\onlinecite{PhysRevB.86.241111} that for a pyrochlore slab, the underlying stacking arrangement leads to intriguing band structures with relatively high Chern numbers. 

 Recently, there has been a renewal of interest in analyzing the topological features of a closely related cousin of the Kagome lattice. This is called the Lieb lattice (which is an example of a depleted-square lattice with space group $p4mm$); it can also be obtained by continuously shearing an ideal Kagome structure.  The continuous evolution of the band structure, including the flat band and its topological variation between the Lieb and the Kagome lattice, has been recently studied in Refs.~\onlinecite{PhysRevB.99.125131,PhysRevB.101.045131}. This lattice was previously studied~\cite{Tsai_2015} in detail for its topologically protected quadratic band crossing point (QBCP) in the band structure. QBCP \--- a Brillouin zone (BZ) point where two bands cross each other with a quadratic dispersion \---  is a generic feature of certain type of lattices, where discrete crystal symmetries play an important role~\cite{PhysRevLett.103.046811}. However, unlike its linear band crossing counterpart, namely the Dirac point, a QBCP is not robust under an arbitrarily weak interaction~\cite{Wehling,Banerjee_2020,PhysRevB.77.235125}. Here, we note that Lieb lattices have been experimentally realized recently in photonic~\cite{PhysRevB.81.041410,Guzm_n_Silva_2014,PhysRevLett.114.245504,PhysRevLett.114.245503,Xia:16,PhysRevLett.116.183902}, electronic~\cite{Slot2017} and cold-atom settings~\cite{Taiee1500854}.

 In this paper, we integrate these research directions and consider a tight-binding model of intrinsically spin-orbit coupled electrons on quasi-two dimensional systems composed of stacked multilayer Lieb lattices. We show that for two distinct Bernal-type stackings the previously known single-layer QBCPs evolve into {\it extended degeneracy lines} along the Brillouin zone (BZ) edge \--- which we designate as  Quadratic Band Crossing Lines (QBCL). We further motivate that the QBCLs are the generalization of QBCPs for non-symmorphic lattice structures, and are protected by discrete symmetries. In the multilayer set-up with spin-orbit coupling, these QBCLs form well-separated band-subspaces. Their topological features are computed by analyzing the associated Wilson spectrum, from which we obtain the  corresponding Chern numbers. Then, we discuss one of our {\it main results} that the layer number $\mathcal{N}$ in the multilayer structure offers natural tunability to generate an emergent band structure with higher Chern numbers, and as such, provides a unique platform for exploring hitherto unknown topological phases of matter.

%================================================================================================================================
\section{Monolayer Lieb lattice \label{sec:mono_lieb}}
%================================================================================================================================

We start from an extended tight-binding (TB) model with intrinsic spin-orbit coupling (ISOC) in the single-layer Lieb lattice~\cite{PhysRevB.82.085310}, and consider stacking them in a multilayer structure. The single-layer Hamiltonian is written as
\begin{equation}\label{eq.1}
\mathcal{H}_{\text{sl}} = \sum_{i\sigma} \varepsilon_{i} c^{\dagger}_{i\sigma}c_{i\sigma}-\sum_{ij\sigma} t_{ij} c^{\dagger}_{i\sigma}c_{j\sigma} + i\lambda \sum_{\llangle ij \rrangle} \nu_{ij} c^{\dagger}_{i\alpha} \sigma^z_{\alpha \beta} c_{j\beta},
\end{equation}
where $c^{\dagger}_{i\sigma}$ creates an electron at site $i$ with spin $\sigma$ and $\sigma_z$ is the Pauli matrix. In the last
%=================================================================================================================================
\begin{figure}[t]
\centering
\includegraphics[width=1\linewidth]{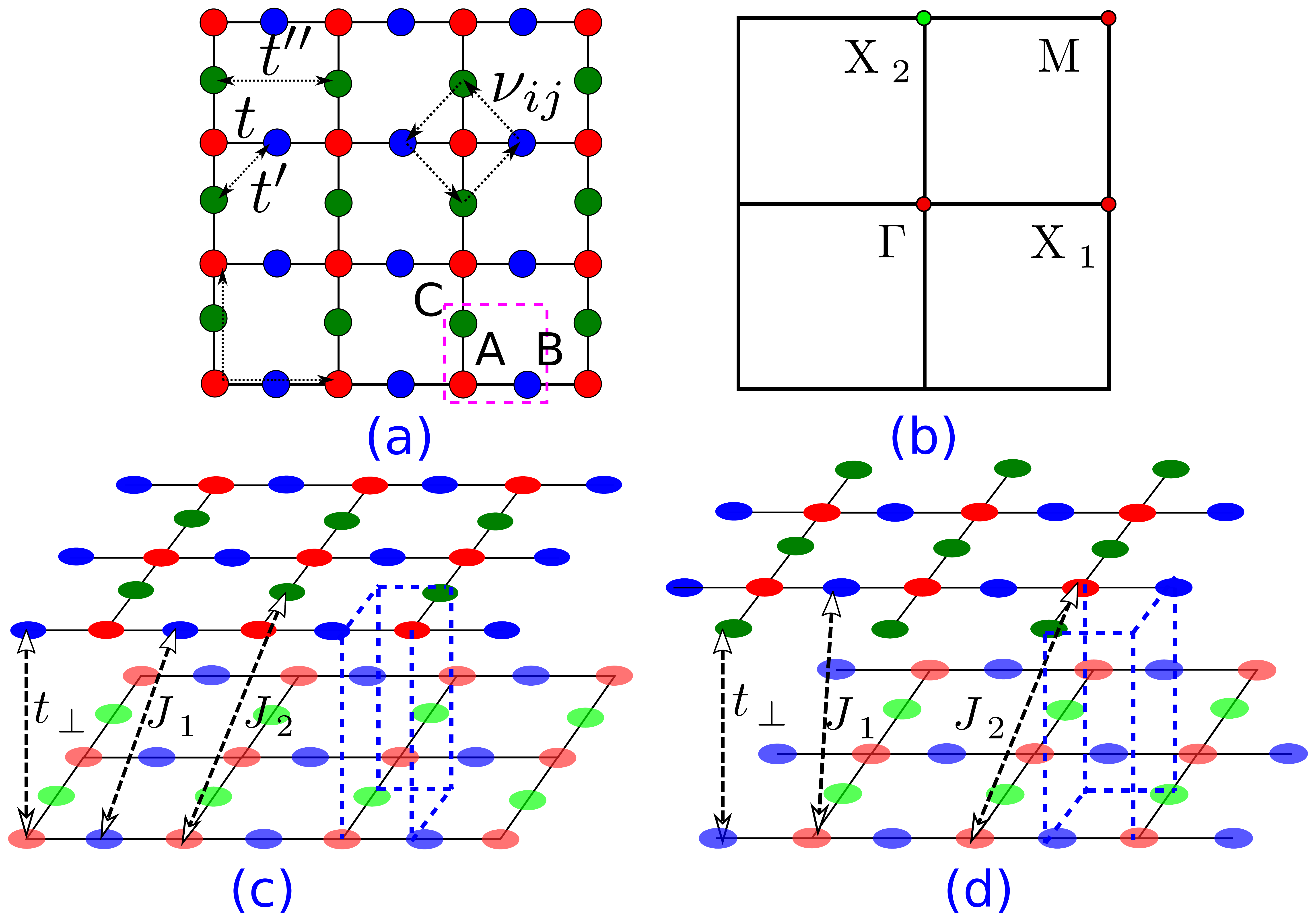} 
\caption{(a) A schematic of the two-dimensional Lieb lattice with all possible hoppings $t_{ij}$ between the sites. The arrows corresponding to $\nu_{ij}$ show the unit vectors related to the intrinsic spin-orbit coupling and are discussed in the text. The three-atom unit cell is marked in a red-dashed line. (b) The Brillouin zone for the bilayer system with the high-symmetry points illustrated by  filled circles. (c,d) A pictorial representation of a bilayer coupled Lieb lattice arranged in two different stackings \--- AB and ABC, respectively. The blue-dashed region signifies the modified unit cell when a multilayer structure is incorporated. }
\label{fig:Fig1}
\end{figure} 
%=================================================================================================================================
term, we assume summation over the repeated indices. The hopping amplitude $t_{ij}$ is considered finite between the first ($t$), the second ($t'$) and the third ($t''$) nearest-neighbor sites, and $\varepsilon_i$ labels the onsite energies for the three sublattice sites (see Fig.~\ref{fig:Fig1}a). In general, the three onsite energies can be different. Yet, the four-fold rotation symmetry enforces the edge-centered site energies to be equal \textit{i.e.} $\varepsilon_B=\varepsilon_C$. Therefore without loss of generality, we assume a finite $\varepsilon_A$ with vanishing $\varepsilon_B,\varepsilon_C$. The longest hopping amplitude $t''$ is considered only when there is no site in between the relevant hopping process~\cite{Tsai_2015}. Finally, $\lambda$ is the strength of the ISOC between the next-nearest neighbor sites, and $\nu_{ij} = \bm{\hat{d}}^1_{ij}\times \bm{\hat{d}}^2_{ij} = \pm 1$. Here $\bm{\hat{d}}^1_{ij}$ and $\bm{\hat{d}}^2_{ij}$ denote the two unit vectors connecting the second neighbor sites, as illustrated in Fig.~\ref{fig:Fig1}a. For an explicit construction of $\mathbf{H}_{\text{sl}}$, and its associated band structure, we refer to Appendix~\ref{sec:app1}. For completeness, we also add a brief discussion on the effect of Rashba spin-orbit coupling (spin non-conserving part) in Appendix~\ref{sec:app2}.

The above Hamiltonian can be written in Fourier space as $\mathcal{H}_{\text{sl}} = \sum_{\vk} \Psi^{\dagger}_\vk \left(\mathbb{I}_2 \otimes \mathbf{H}_{\vk,\text{sl}} \right) \Psi_\vk$, where the spinor $\Psi_{\vk}$ is composed of three operators $c_{\vk,\alpha,\sigma}$ on the three sublattices $\alpha = A,B,C$, with spin projection $\sigma$. We notice that the Hamiltonian is composed of two uncoupled spin-projected Hamiltonians $\mathbf{H}_{\vk,\text{sl}}$. The spin-degeneracy is not broken as a result of the inversion symmetric ISOC. Therefore for subsequent analysis, we focus on only one of the spin-projected Hamiltonians. 

%=================================================================================================================================
\begin{figure}[t]
\centering
\includegraphics[width=1\linewidth]{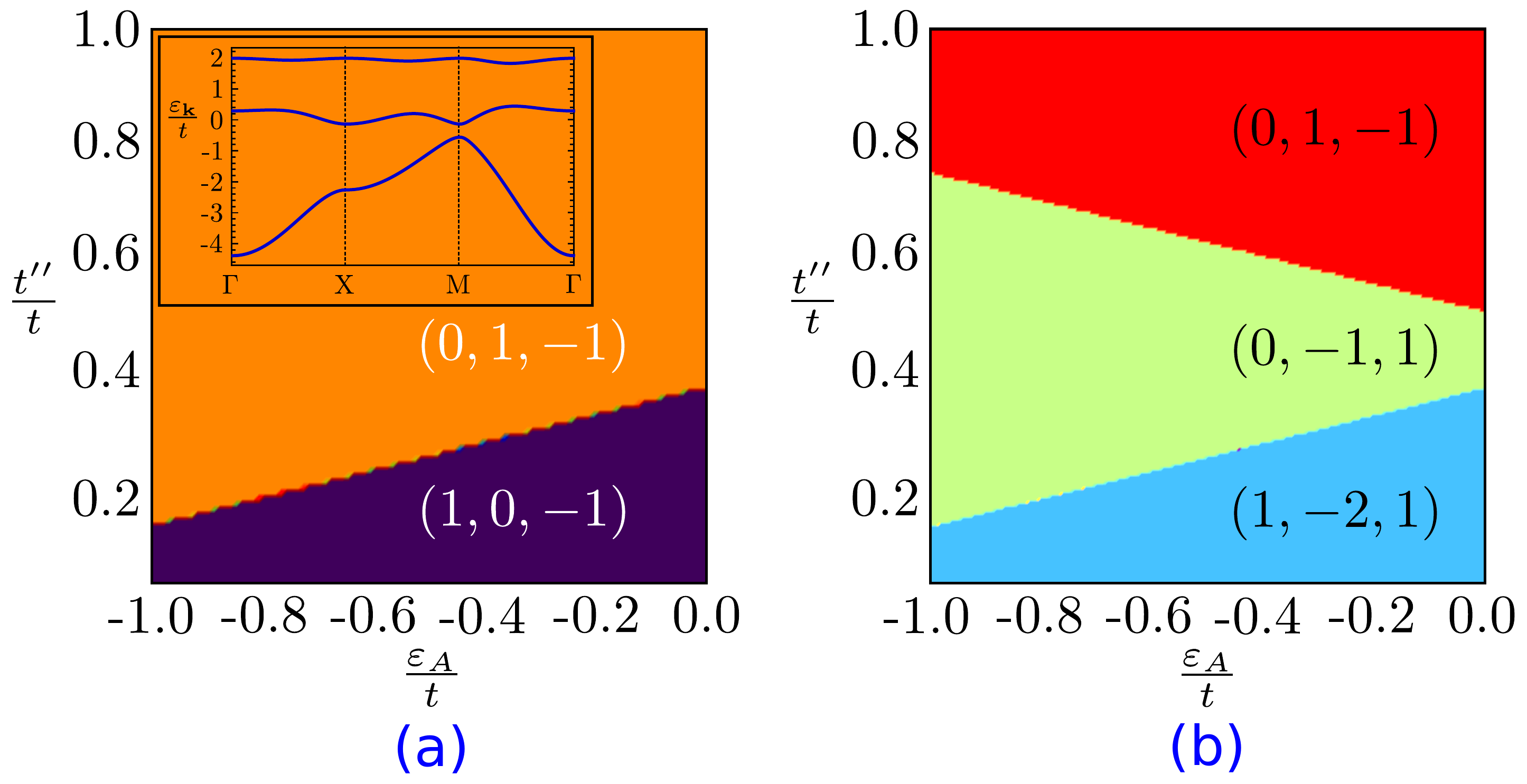} 
\caption{Topological phase diagram in the single-layer Lieb lattice as a function of $t''$ and $\varepsilon_A$ in the presence of intrinsic spin-orbit coupling $\lambda = 0.35 t$: (a) in the case of the next-neighbor hopping $t'(=0.3t) < 0.5 t$ and (b) $t' (= 0.75t) >0.5 t$. The intersection between the different colored areas indicates the closing of one of the band gaps. The Chern number distribution for spin-up bands is arranged from the lowest to the highest bands as shown in the inset of panel (a). Inset: The band structure with finite intrinsic SOC with the tight-binding parameters $t=1, t'=0.3,t''=0.2, \varepsilon_A = -1, \lambda = 0.35$. All values are in the units of eV. }
\label{fig:Fig2}
\end{figure} 
%=================================================================================================================================  

The ISOC incorporates non-trivial topological character in the band structure for the single-layer Lieb lattice. For a better understanding, we compute the topological Chern number for each band. An example of the gapped band-structure is shown in the inset of Fig.~\ref{fig:Fig2}a. For a detailed evolution of the ISOC band-structure with various TB parameters, we refer to Appendix~\ref{sec:app1}. The Chern number for a particular isolated band $\varepsilon^n_\vk$ is defined as $\mathcal{C}_n = \frac{1}{2\pi}\int_{\text{BZ}}\mathsf{\Omega}_n(\vk)d\vk$, where $\mathsf{\Omega}_n(\vk)=\partial_x \mathcal{A}^n_y(\vk)-\partial_y \mathcal{A}^n_x(\vk)$ is the Berry curvature. Here, $\mathcal{A}^n_i(\vk) = i\Braket{\psi^n_\vk|\partial_i|\psi^n_\vk}$ is the Berry connection for the corresponding band $\varepsilon^n_\vk$, with eigenfunctions $\ket{\psi_\vk^n}$. In this paper, we numerically compute the Chern numbers using the method of link variables~\cite{Fukui}, and obtain the Chern number distribution for the gapped bands (arranged from the bottom to the top-most band) as $\mathcal{C} = (1,0,-1)$ in the absence of $\varepsilon_A, t''$ and, for $t'/t < 0.5$. The topological evolution of such Chern number distribution as a function of $t'$ and $\lambda$ has been analyzed in detail in an earlier theoretical work~\cite{PhysRevB.86.195129}.

%================================================================================================================================
\subsection{Chern number distribution \label{sec:chn_dist}}
%================================================================================================================================

Here, we identify that the onsite energy $\varepsilon_A$ and the longest hopping $t''$ produce an {\it even richer phase diagram} with versatile topological characteristics. We notice that for $t'<0.5t$, the three bands from bottom to top have a distribution $\mathcal{C} = (1,0,-1)$ for small $t''$. It eventually changes to $\mathcal{C} = (0,1,-1)$ for larger $t''$ as shown in Fig.~\ref{fig:Fig2}a. However, for $t'>0.5t$ we notice three different topological phases. For small enough $t''$ and $\varepsilon_A$, the distribution is $\mathcal{C} = (1,-2,1)$.  It changes to $(0,-1,1)$ for an intermediate $t''$ and eventually becomes $(0,1,-1)$ for sufficiently large $t''$ (see Fig.~\ref{fig:Fig2}b). The interface between the colored regions (phases) in Fig.~\ref{fig:Fig2} indicates the absence of gaps in the band structure. For this analysis, we fixed the magnitude of the ISOC as $\lambda = 0.35 t$. Apart from quantitative changes in the overall size of the individual phases in Fig.~\ref{fig:Fig2}, our analysis remains qualitatively valid for arbitrary $\lambda$, provided all the three bands are fully gapped.

%================================================================================================================================
\section{Stacked Lieb lattices \label{sec:stack_lieb}}
%================================================================================================================================

We now focus on the main part of this work, and discuss the implications for the stacked Lieb layers. For the subsequent analysis, we primarily consider two Bernal-type stackings: AB and ABC. A discussion of the conventional AA-stacking is provided in Appendix~\ref{sec:app3}. For a pictorial illustration of the bilayer setup, see Fig.~\ref{fig:Fig1}c,d. In AB-stacking, the top layer is shifted by a half-lattice vector ($\bm{a}_1/2$) along the horizontal direction with respect to the bottom layer, whereas in ABC-stacking it is shifted by the same amount along both the axial directions ($\bm{a}_1/2, \bm{a}_2/2$). This fractional translation in the Bernal-stacked configurations leads to an {\it emergent non-symmorphic} crystal structure, even though the single-layer Lieb lattice is characterized by a symmorphic space group $p4mm$. A fundamental difference between these two symmetries stems from how the spatial origin evolves under the allowed transformations: symmorphic symmetries preserve the origin, while non-symmorphic symmetries lead to a fractional shift of the origin~\cite{PhysRevX.6.021008}. Specifically, the layer groups associated with AB- and ABC-stacking are $pmma$ and $p4/nmm$, respectively~\cite{Hitzer2013,PhysRevB.101.165121,PhysRevLett.115.126803}.

The inter-layer couplings are assumed to be $(t_{\perp},J_1,J_2)$ as illustrated in Fig.~\ref{fig:Fig1}c,d. The unit-cell consists of two dimer and two monomer sites (dashed lines in Fig.~\ref{fig:Fig1}c,d). The dominant interlayer coupling $t_{\perp}$ is considered between the atoms in each layer within the dimer site, whereas the remote hoppings $J_1$, $J_2$ are considered between the dimer and the monomer sites (as illustrated in Fig.~\ref{fig:Fig1}c,d). Consequently, the Hamiltonian for the coupled system in terms of a spinor $\Psi_\vk = (c_{1\text{A}\vk},c_{1\text{B}\vk},c_{1\text{C}\vk},c_{2\text{A}\vk},c_{2\text{B}\vk},c_{2\text{C}\vk})$ is written as
\begin{equation}\label{eq.2}
\mathbf{H}_{\vk,\text{ab/abc}} = \mathbf{H}_{\vk,\text{sl}}\otimes \mathbb{I}_{\sigma} + \mathbf{V}_{\vk,\text{ab/abc}}\otimes \sigma_1,
\end{equation}
where $\bm{\sigma}$ are the Pauli matrices representing the layer degrees of freedom, $\mathbb{I}_{\sigma}$ is a $2\times 2$ identity matrix, $\mathbf{H}_{\vk,\text{sl}}$ is the single-layer TB Hamiltonian defined in Eq.~\ref{eq.1}, and $\mathbf{V}_{\vk,\text{ab/abc}}$ corresponds to the interlayer coupling Hamiltonian for the two stackings defined  as%=================================================================================================================================
\begin{figure}[t]
\centering
\includegraphics[width=1\linewidth]{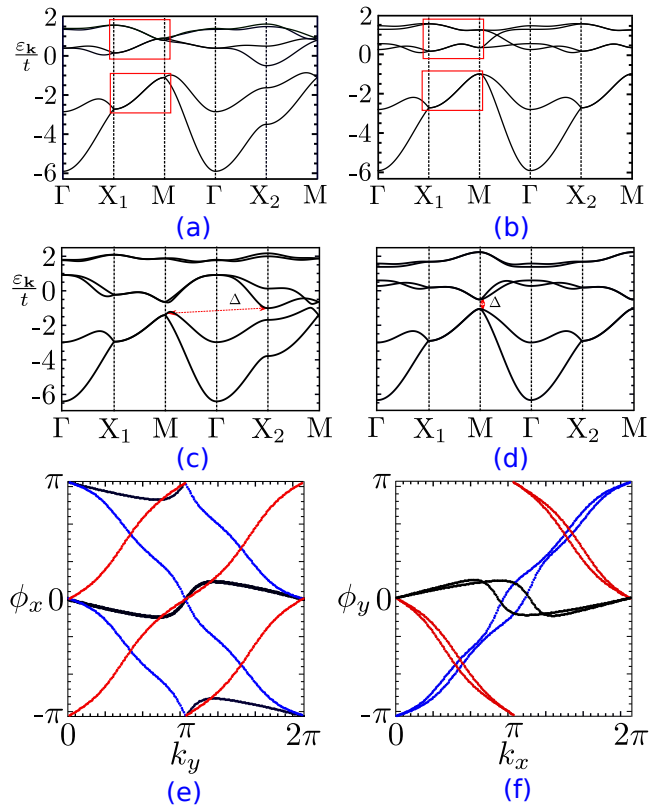} 
\caption{(a,b) The band structure, along the high-symmetry directions in the BZ for AB- and ABC-stacked bilayer Lieb lattices, respectively. The TB parameters for each layer are chosen as $t=1, t'=0.3,t''=0.2, \varepsilon_A = -1$ with the interlayer couplings $t_{\perp} = 0.45,J_1=0.25, J_2 = 0.15$. All values are in the units of eV. (c,d) The gapped spectrum with gap $\Delta$ in the presence of intra-layer intrinsic spin-orbit coupling $\lambda = 0.35$. The other parameters remain identical as in panels (a,b). (e,f) The non-abelian Berry phases for the three two-band subspaces obtained from the Wilson loop (Eq.~\ref{eq.4}) for the AB-stacked bilayer Lieb lattice with $\lambda = 0.35$. The apparently reduced $C_4$ rotation in the AB-stacking leads to two inequivalent Wilson spectra along the two axial directions in the BZ. For the ABC-stacking the Wilson spectra (not shown here) along both axial directions are equivalent and they are identical to panel (e). (Color scheme: black - lowest QBCL, blue - middle QBCL and red - top QBCL).}
\label{fig:Fig3} 
\end{figure} 
%=================================================================================================================================  
\begin{subequations} \label{eq.3}
   \begin{align} 
    \mathbf{V}_{\vk,\text{ab}} &= -2J_1c_1\mathbb{I}_3 -t_{\perp} \mathsf{\Gamma}_1 -4J_2c_1c_2 \mathsf{\Gamma}_4 - 2J_1 c_2 \mathsf{\Gamma}_6 \label{eq.3a} \,, \\
    \mathbf{V}_{\vk,\text{abc}} & = -4J_2c_1c_2\mathbb{I}_3 -2J_1c_2 \mathsf{\Gamma}_1  -2J_1c_1\mathsf{\Gamma}_4 - t_{\perp} \mathsf{\Gamma}_6 \label{eq.3b} \,, \\
  \mathsf{\Gamma}_1  =& 
  \begin{pmatrix}
  0 & 1 & 0 \\
  1 & 0 & 0 \\
  0 & 0 & 0
  \end{pmatrix}, ~~
\mathsf{\Gamma}_4  =
 \begin{pmatrix}
  0 & 0 & 1 \\
  0 & 0 & 0 \\
  1 & 0 & 0
 \end{pmatrix}, ~~
\mathsf{\Gamma}_6  =
 \begin{pmatrix}
  0 & 0 & 0 \\
  0 & 0 & 1 \\
  0 & 1 & 0
 \end{pmatrix}, 
\end{align} 
\end{subequations}
where $c_{i}= \cos k_i/2$, $k_i = \vk\cdot \bm{a}_i$, and $\mathsf{\Gamma}_i$'s are the traceless Gell-Mann matrices. The energy spectrum is obtained by diagonalizing the Hamiltonian $\mathbf{H}_{\vk,\text{ab/abc}}$. The corresponding band structures for both stackings are shown in Fig.~\ref{fig:Fig3}a-d. %, respectively.
 The top panel in Fig.~\ref{fig:Fig3} shows the dispersion without the ISOC, whereas the middle panel displays the gapped spectrum with intra-layer ISOC. Here, the left and the right panels correspond to AB- and ABC-stackings, respectively.  Interestingly, the gapped band structures in (c) and (d) with gap $\Delta$ are analogous to the indirect- and direct-gap semiconductors, respectively.

%================================================================================================================================
\subsection{Quadratic Band Crossing Lines \label{sec:qbcl}}
%================================================================================================================================

The emergent non-symmorphicity leads to a strikingly different feature in the resulting band structure \--- three pairs of bands individually become degenerate along the extended region of the BZ edge. In the case of AB-stacking, the directions X$_1 \rightarrow$ M and X$_2 \rightarrow$ M become inequivalent as shown in Fig.~\ref{fig:Fig3}a,c, due to the partially broken $C_4$ rotation symmetry. It leads to band degeneracy only along X$_1 \rightarrow$ M whereas the degeneracy is lifted along X$_2 \rightarrow$ M. In contrast, for ABC-stacking the band degeneracy exists along the entire perimeter of the BZ as shown in Fig.~\ref{fig:Fig3}b,d. To further understand the structure of these {\it degenerate band lines}, we first focus on the lowest two bands in Fig.~\ref{fig:Fig3}a,b without the ISOC, as they remain well separated from the other bands. Consequently, we consider the quasiparticle dynamics near $1/6$ filling (for the spin-polarized case). In this case, the quasiparticles will mostly populate the two $A$ sites on each layer. To obtain an effective Hamiltonian, we expand the $6\times6$ Hamiltonian in Eq.~\ref{eq.2} close to the M point and integrate out the other degrees of freedom~\cite{PhysRevB.101.045131} to obtain
\begin{subequations}
\begin{align} 
\mathbf{H}_{\vk,\text{ab}}^{\text{eff}} & \approx \left(\varepsilon_A -\frac{t_{\perp}^2+t^2 \vk^2}{2t''-\varepsilon_{\mathrm{F}}}\right)\cdot \mathbb{I}_2 + \frac{2t_{\perp}tk_1}{2t''-\varepsilon_{\mathrm{F}}}\cdot \sigma_1 \label{eq.effective_ab}
\,, \\
\mathbf{H}_{\vk,\text{abc}}^{\text{eff}} & \approx \Big[\varepsilon_A + \frac{t^2(2t''-\varepsilon_{\mathrm{F}})\vk^2}{\mathcal{S}}\Big]\cdot \mathbb{I}_2 + \frac{2t_{\perp}t^2k_1k_2}{\mathcal{S}}\cdot \sigma_1 \,, \label{eq.effective_abc}
\end{align}
\end{subequations} 
where $\mathcal{S} = t_{\perp}^2-4t''^2+4t''\varepsilon_{\mathrm{F}}-\varepsilon_{\mathrm{F}}^2$ and $\varepsilon_{\mathrm{F}}$ is the chemical potential at $1/6$ filling. Here, $\sigma_1$ is the Pauli matrix corresponding to the layer degrees of freedom. For simplicity, we ignored the long-range interlayer hopping amplitudes $J_{1},J_2$, which will further renormalize the Fermi velocity. For both the stackings, we obtain quadratic dispersion along M $\rightarrow$ X$_1$ following Eq.~\ref{eq.effective_ab},\ref{eq.effective_abc}. A similar analysis of the effective Hamiltonian around X$_1$ yields analogous quadratic dispersions for these band crossing lines in both AB- and ABC-stacking. Here, we emphasize that the corresponding dispersions of course do not remain quadratic all along the BZ edge, as is evident from Fig.~\ref{fig:Fig3}a,b. However, because of the asymptotic behavior of the band crossing lines at X and M points, we designate them as quadratic band crossing lines (QBCL).

Next, we analyze the effects of ISOC on the double layer system and explore the possibility of any {\it topological transitions}. The degenerate QBCL structure in the spectrum is preserved in both AB- and ABC-stacked bilayer systems (see Fig.~\ref{fig:Fig3}c,d), even in the presence of ISOC.  We envisage that the origin of this degeneracy for the QBCLs is tied to the fractional glide transformations $\{\mathsf{g}_x|(\frac{1}{2},0)\}$ and $\{\mathsf{g}_y|(0,\frac{1}{2})\}$ of the underlying layer groups. For perpendicular axial glide transformation, the QBCL degeneracy remains intact (see Appendix~\ref{sec:app4} for details)~\cite{Schoop2016,PhysRevLett.115.126803}. Here, $(\frac{1}{2},0),(0,\frac{1}{2})$ represent the half-translations along respective crystal directions. We observe that within a finite region of the parameter space in our model, the three different QBCLs form two-band subspaces well separated by the two bandgaps of almost equal magnitude $\Delta$ (see Fig.~\ref{fig:Fig3}c,d).

%================================================================================================================================
\subsection{Wilson loop computation \label{sec:wilson}}
%================================================================================================================================

As the spectra in Fig.~\ref{fig:Fig3}a-d contain non-separable bands, we utilize a different scheme (from the monolayer Lieb lattice case) to compute the Chern number, by analyzing the {\it multi-band non-abelian} Berry phases~\cite{PhysRevB.89.155114,Lu2016,PhysRevB.100.195135,PhysRevB.95.241101,vanderbilt_2018} for each of the two-band subspaces. The Chern number is then computed from the non-trivial windings of these Berry phases. The latter are computed from the overlap matrices $F_{mn}(\vk;\vk+\Delta \vk) = \Braket{u_m(\vk)|u_n(\vk+\Delta \vk)}$, where $\ket{u_{m}(\vk)}$ are the Bloch wave-functions obtained by diagonalizing the Hamiltonian in Eq.~\ref{eq.2}. We multiply these overlap matrices to construct the Wilson loop operator 
\begin{equation}\label{eq.4}
\mathbf{W}_{k_{\alpha}} = \overline{\prod_{k_{\beta}}}\mathbf{F}(k_{\beta}|k_{\beta}+\Delta k_{\beta}), \quad \alpha,\beta = x,y,
\end{equation}
where $\mathbf{F}(k_{\beta}|k_{\beta}+\Delta k_{\beta})$ is the $2\times 2$ matrix composed of $F_{mn}$ for each of the two-band subspaces, and $\overline{\Pi}$ implies path-ordered product of the overlap matrices along a closed loop in the two-dimensional BZ, \textit{i.e.} for $k_{\beta}$ ranging between  $0$ and $2\pi$. For the purpose of this work, we consider two different loops in the BZ: (i) by fixing $k_x$, we consider a Wilson loop along $k_y$ between $0$ and $2\pi$, and (ii) for fixed $k_y$, the loop is considered from $k_x = 0$ to $k_x = 2\pi$. Because the discretization of the BZ incorporates non-unitary effects on the overlap matrices, one needs to fix a gauge while performing the numerical computation. Here, we implement a periodic gauge at the two ends of the respective Wilson loop, \textit{i.e.} at $k_{\beta} = 0$ and $k_{\beta} = 2\pi$. We set $\ket{u_m(\vk_0 +\mathbf{G})} = e^{-i\mathbf{G}\cdot \mathbf{r}} \ket{u_m(\vk_0)}$, where $k_0 = 0$ and $\mathbf{G}$ corresponds to the  reciprocal lattice vector. For the rest of the $\vk$-points in the loop, such a gauge fixing is not required~\cite{Lu2016}. The Berry phases $\phi_{\vk_{\perp}}$ are then computed from the eigenvalues $\lambda_{\vk_{\parallel}}$ of the Wilson loop operator $\mathbf{W}_{k_{\parallel}}$, as $\phi_{\vk_{\parallel}} = - \text{Im} \log{\lambda_{\vk_{\parallel}}}$.

The non-abelian Berry phase spectra (along both the axial directions in the BZ) for AB-stacked bilayer Lieb lattice with ISOC are shown in Fig.~\ref{fig:Fig3}e,f. For the lowest QBCL (black curve), we notice the Berry phase does not wind at all between $\pi$ and $-\pi$, and hence the Chern number for the lowest two-band subspace is $0$. However, the middle (blue curve) and top (red curve) QBCLs along with their respective two-band subspaces contain a non-zero Chern number, as the respective Berry phases exhibit non-trivial winding. By counting the winding number, we find that the middle and top QBCL acquire Chern numbers $2$ and $-2$, respectively. Note that the horizontal glide in AB-stacking reduces the $C_4$ rotation symmetry, and hence the Wilson spectra along the two axial directions in the BZ become inequivalent as contrasted in Fig.~\ref{fig:Fig3}e and Fig.~\ref{fig:Fig3}f. In comparison, for ABC-stacking the glide is applied in both the axial directions and hence the Wilson spectra (not shown here) are equivalent along both the directions and are identical to $\mathbf{W}_{k_x}$ eigenvalues for AB-stacked case (Fig.~\ref{fig:Fig3}e). However, the Chern number distribution (bottom to top QBCLs) remains the same as $(0,2,-2)$. Since, the analysis is done for one spin component (see the monolayer discussion), we can compute the spin Chern number $\mathcal{C}_{\text{spin}}$ as $\mathcal{C}_{\text{spin}} = \mathcal{C}_{\uparrow} - \mathcal{C}_{\downarrow}$. As the Chern number for spin-up and spin-down components differs in sign for a time-reversal symmetric system~\cite{PhysRevB.86.195129,PhysRevLett.97.036808}, we obtain $\mathcal{C}_{\text{spin}} = (0,4,-4)$, which has implications for the spin-Hall conductivity $\sigma_{\text{sh}} = e^2/\hbar \sum_{\varepsilon < \varepsilon_{\mathrm{F}}} \mathcal{C}_{\text{spin}}$~\cite{Dayi.2016}. 

%=================================================================================================================================
\begin{figure}[t]
\centering
\includegraphics[width=1\linewidth]{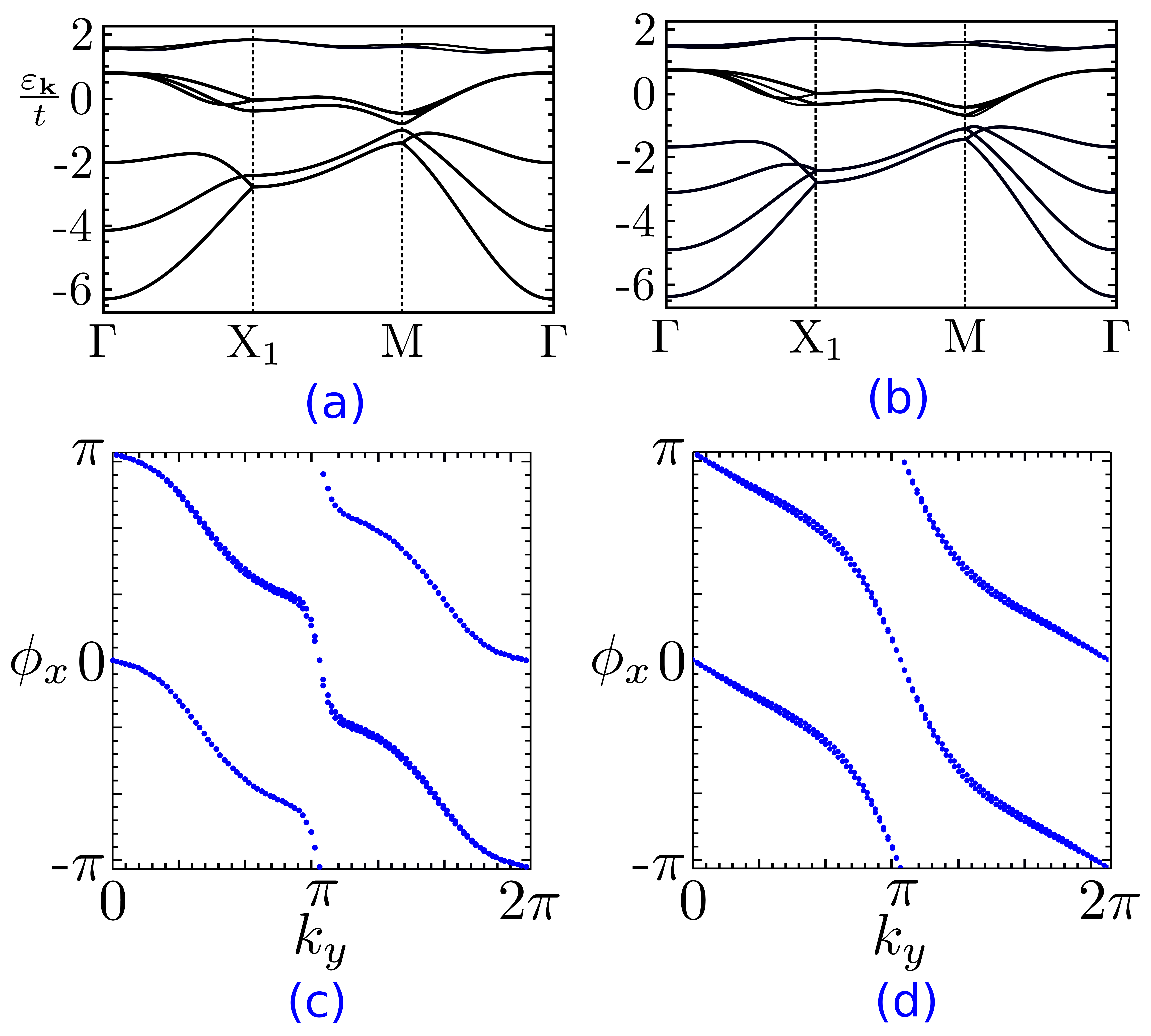} 
\caption{The spectrum for trilayer (a) and quadrilayer (b) AB-stacked Lieb lattice with the emergent QBCLs in the presence of intrinsic spin-orbit coupling in each layer. The associated non-abelian Berry phase for the \textit{middle band subspace} computed from the Wilson loop operator for the trilayer (c) and the quadrilayer (d), respectively.}
\label{fig:Fig4} 
\end{figure} 
%================================================================================================================================

%================================================================================================================================
\subsection{Multilayer Stacking \label{sec:mult_stck}}
%================================================================================================================================

Motivated by the observation of Chern number $2$ for the bilayer stacking, we  now consider a multilayer generalization. Considering the two basic stackings (AB and ABC), the number of possible orientations for an $\mathcal{N}$-layer system grows exponentially as $2^{\mathcal{N}-1}$. However, for simplicity we always keep the stacking between any two adjacent layers as either AB- or ABC-type. The interlayer couplings ($t_{\perp}, \; J_1, \; J_2$ as in Fig.~\ref{fig:Fig1}c,d) are considered only between the two adjacent layers. Again within a finite region of our parameter space, we notice that the spectrum is divided into three band-subspaces well separated from each other. For an $\mathcal{N}$-layer system, each band-subspace contains $\mathcal{N}$-bands, with $\floor{\frac{\mathcal{N}}{2}}$ QBCLs along the BZ edge. Consequently, for even number of layers the spectrum consists of only QBCLs along the BZ edge (see Fig.~\ref{fig:Fig4}b), while for odd number of layers, each of the three band-subspaces contains one lone band (see Fig.~\ref{fig:Fig4}a) along with $\floor{\frac{\mathcal{N}}{2}}$ QBCLs. 

The other properties of spectrum simply follow from our analysis of the bilayer Lieb system, and hold true for the multilayer setup as well. The topological character of the gapped bands are again analyzed with the Wilson loop technique. Consequently, we compute the non-abelian Berry phases and find that for the $\mathcal{N}$-layer system, the Chern number distribution is arranged as $\mathcal{C} = (0,\mathcal{N},-\mathcal{N})$. An illustration of the band structure and the associated non-abelian Berry phases (for the middle band subspace) is shown in Fig.~\ref{fig:Fig4}a,b for the AB-stacked tri-(odd) and quadri-(even)-layer setups. We have explicitly checked the validity of this result for the number of layers up to 10 (see Appendix~\ref{sec:app5} for more details). Hence we propose it to be a generic feature of the non-symmorphic Lieb multilayers. The layer number $\mathcal{N}$ naturally offers tunability to the topological Chern number, and hence is measurable in the spin-Hall conductivity $\sigma_{\text{sh}}$. 

%================================================================================================================================
\section{Discussion and Conclusion \label{sec:conclu}}
%================================================================================================================================

The Lieb lattice is unique in that it provides an ideal depleted lattice in two dimensions and also has its depleted three-dimensional analog. Moreover, it is the sheared limit of the Kagome lattice when the $120^{\circ}$ angle becomes $90^{\circ}$.  Interestingly, the Kagome lattice is maximally frustrated whereas the Lieb lattice is unfrustrated when one considers spin phenomena on such lattices.  For in-between shear angles, there is an intermediate Lieb-Kagome (or depleted oblique) lattice \cite{PhysRevB.99.125131} which interpolates between the two limiting lattices and is of interest in its own right. The stacked Lieb lattice provides an even more elaborate platform for exploring novel topological phenomena, phases, and transitions \--- QBCLs and {\it higher} Chern numbers being two such examples.  In doing so we had to generalize the Wilson loop method to QBCLs, which is a {\it tour de force} technique for extended band degeneracies and can be adopted in a wide variety of physical contexts.  

In conclusion, our main findings are as follows: (i) Bilayer Lieb lattice provides a natural harbor for hosting QBCLs, (ii) QBCLs are a generalization of QBCPs, and (iii) non-symmorphicity is a necessary condition for QBCLs. To calculate non-abelian Berry phases and Chern numbers around QBCLs we devised a powerful, new Wilson loop method computationally~\cite{Abigail}. (iv) We found higher Chern numbers in the band subspace and (v) novel topological transitions including phases involving higher Chern numbers. (vi) The multilayer Lieb lattice band structure is labeled by Chern numbers that are proportional to the number of layers.  Given that Lieb lattices have been experimentally realized recently in photonic, electronic and atomic settings~\cite{PhysRevB.81.041410,Guzm_n_Silva_2014,PhysRevLett.114.245504,PhysRevLett.114.245503,Taiee1500854,Xia:16,PhysRevLett.116.183902,Slot2017}, with the possible fabrication of bilayer Lieb lattice our results indicate that unique topological signatures such as spin Chern numbers, associated spectral functions, etc. can be measured in realistic materials. Recently, a bilayer Lieb lattice system has been fabricated on an acoustic crystal~\cite{Deng2020} and has been shown to possess similar QBCL like features. Finally, we mention that Mielke, and T$_3$ are among the possible other lattices \cite{PhysRevA.99.053608,PhysRevA.80.063603}, where the physics discussed here can also be realized. It would also be interesting to analyze a bosonic analog of our system~\cite{Deng2020}.

\begin{acknowledgements}
We acknowledge helpful discussions with D. Gresch, A. Bouhon, W.~A. Atkinson, D. Vollhardt, and A. P. Kampf. This work was supported in part by the Deutsche Forschungsgemeinschaft (DFG, German Research Foundation)-TRR 80 and in part by the U.S. Department of Energy.   
\end{acknowledgements}

%%%%%%%%%%%%%%%%%%%%%%%%%%%%%%%%%%%%%%%%%%%%%%%%%%%%%%%%%%%%%%%%%%%%%%%%%%%%%%%%%%%%%%%%%%%%%%%%%%%%%%%%%%%%%%%%%%%%%%%%%%%%%%%%
\appendix
%=========================================================================
\section{Single-layer Lieb Lattice: Intrinsic spin-orbit coupling \label{sec:app1}}
%=========================================================================

%=================================================================================================================================
\begin{figure*}[t]
\centering
\includegraphics[width=0.85\linewidth]{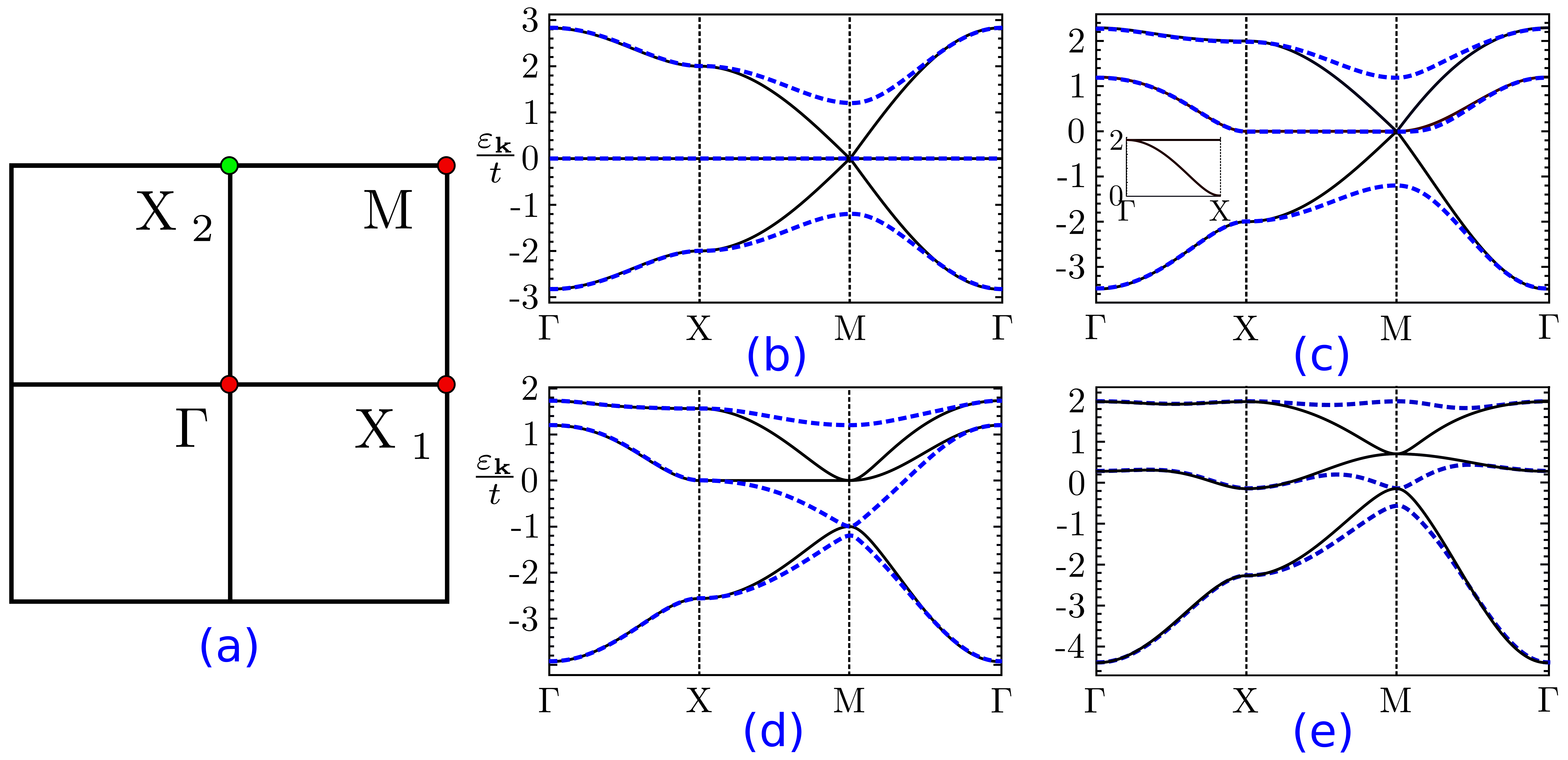} 
\caption{(a) The square-shaped Brillouin zone for the two-dimensional Lieb lattice with high-symmetry points labeled by the filled circles. (b-e) A comparison between the band structures of Lieb lattice with (dashed lines) and without (solid lines) the spin-orbit coupling along the high-symmetry directions in the Brillouin zone for various TB parameters: (b) the nearest-neighbor hopping $t = 1$ with all the other parameters set to zero, (c) next-neighbor hopping $t'=0.3$ with finite $t=1$ (rest of the parameters are zero), (d) $t=1, t'=0.3,t''=0.0, \varepsilon_A=-1$ and (e) $t=1, t'=0.3,t''=0.2, \varepsilon_A = -1$. The spin-orbit coupling strength is $\lambda = 0.35t$. Inset (c): The middle band touches the top band at $\Gamma$ point for $t' = 0.5 t$. All values are in the units of eV.}
\label{fig:AFig1}
\end{figure*} 
%=================================================================================================================================  

In this section, we analyze the evolution of the Lieb lattice band-structure with various tight-binding (TB) parameters as incorporated in Eq. (1) in the main text. The Hamiltonian is rewritten as
\begin{widetext}
\begin{equation}\label{seq.1}
\mathcal{H}_{\text{sl}} = \underbrace{\sum_{i\sigma} \varepsilon_{i} c^{\dagger}_{i\sigma}c_{i\sigma}-t\sum_{\langle ij \rangle \sigma} c^{\dagger}_{i\sigma}c_{j\sigma} -t'\sum_{\llangle ij \rrangle \sigma} c^{\dagger}_{i\sigma}c_{j\sigma} - t''\sum_{\langle\llangle ij \rrangle\rangle \sigma} c^{\dagger}_{i\sigma}c_{j\sigma}}_{\mathcal{H}_0} + i\lambda \sum_{\llangle ij \rrangle} \nu_{ij} c^{\dagger}_{i\alpha} \sigma^z_{\alpha \beta} c_{j\beta},
\end{equation}
\end{widetext}
where the parameters are defined as in the main text. As intrinsic spin-orbit coupling (ISOC) does not break the spin degeneracy, we consider only the spin-up electrons as mentioned in the main text. The Hamiltonian in Eq.~\ref{seq.1} is now rewritten in terms of a three-spinor $\Psi_\vk = (c_{\text{A}\vk},c_{\text{B}\vk},c_{\text{C}\vk})^{\text{T}}$ as 
\begin{widetext}
\begin{align}\label{seq.2}
\mathcal{H}_{\text{sl}} & = \sum_{\vk}\Psi^{\dagger}_{\vk} \mathbf{H}_{\vk} \Psi_{\vk},  ~~\mathbf{H}_{\vk,\text{sl}} = \begin{pmatrix}
\varepsilon_A & -2t\cos \frac{k_1}{2} & -2t\cos \frac{k_2}{2} \\
-2t\cos \frac{k_1}{2} & -2t'' \cos k_2 & -4t'\cos \frac{k_1}{2}\cos \frac{k_2}{2} -4i\lambda \sin \frac{k_1}{2}\sin \frac{k_2}{2} \\
-2t\cos \frac{k_2}{2} & -4t'\cos \frac{k_1}{2} \cos \frac{k_2}{2} +4i\lambda \sin \frac{k_1}{2}\sin \frac{k_2}{2} & -2t'' \cos k_1 \\
\end{pmatrix},
\end{align}
\end{widetext}
where $c^{\dagger}_{\alpha \vk}$ creates an electron on sublattice site $\alpha$ with momentum $k_i = \vk \cdot \mathbf{a}_i$. The unit-vectors are assumed to be $\bm{a}_1 = (a, 0)$ and $\bm{a}_2 = (0, a)$ with $a$ being the lattice constant. The energy spectrum is obtained by diagonalizing the Hamiltonian $\mathbf{H}_{\vk,\text{sl}}$. The corresponding band structures for different TB parameters are shown in Fig.~\ref{fig:AFig1}b-e (solid lines). The ideal Lieb lattice (a finite hopping amplitude $t$ with all other parameters vanishing) has a completely flat and two dispersing bands which cross each other at the M point in the Brillouin Zone (BZ) as shown in Fig.~\ref{fig:AFig1}b. We notice that the complete flatness of the middle band is reduced to a partial one along the BZ edge (X $\rightarrow$ M point) in the presence of the next-nearest neighbor hopping $t'$. With further increasing the strength of $t'$, this middle band becomes more dispersive and eventually touches one of the two other dispersing bands at the $\Gamma$ point at $t'=0.5 t$ as shown in the inset of Fig.~\ref{fig:AFig1}c~\cite{PhysRevB.86.195129}. Whether the middle band touches the top or bottom band, depends on the sign of the hopping parameters. Yet, in both these cases, all the three bands cross each other at the M point~\cite{PhysRevB.86.195129,Tsai_2015}. 

The band crossing at the M point in Fig.~\ref{fig:AFig1}b,c, provides an impression that two of the bands cross each other linearly and there is a Dirac point. However, the structure of the low energy quasiparticles around the M point is different from a Dirac structure. The three-band crossing point is the example of an accidental crossing, which is, indeed, eliminated in the presence of a finite onsite energy $\varepsilon_A$ or second-neighbor hopping $t''$. Finite $\varepsilon_A$ induces a gap at the M point, where the top two bands are separated from the bottom band (see Fig.~\ref{fig:AFig1}d), still retaining the partial flatness of the middle band along the BZ edge. A finite $t''$ completely destroys the flatness, as shown in Fig.~\ref{fig:AFig1}e. However, the band degeneracy at the M point is still preserved (see Fig.~\ref{fig:AFig1}d,e), yielding a quadratic band crossing point (QBCP)~\cite{Tsai_2015}. In the presence of a finite ISOC all the bands are gapped from each other as illustrated by the dashed blue lines in Fig.~\ref{fig:AFig1}b-e. The consequent topological classification is discussed in the main text.

%=========================================================================
\section{Single and Bi-layer Lieb Lattice: Rashba spin-orbit coupling \label{sec:app2}}
%=========================================================================
 
As mentioned in the main text, here we discuss the implication of the Rashba SOC (RSOC) in the monolayer Lieb lattice. For simplicity, we ignore the intrinsic SOC and only focus on the spin non-conserving Rashba effect. The corresponding Hamiltonian is written as
\begin{equation}\label{req.1}
\mathcal{H}_{\mathrm{sl-R}} = \mathcal{H}_0 + i\lambda_{\mathrm{R}} \sum_{\langle ij \rangle \alpha \beta} c^{\dagger}_{i\alpha} \left(\bm{\sigma}\times \bm{\hat{d}}_{ij} \right)_z c_{j\beta},
\end{equation}
where $\lambda_{\mathrm{R}}$ is the strength of the RSOC, and $\bm{\hat{d}}_{ij}$ is the unit vector connecting the nearest-neighbor sites. Since the SOC breaks the spin-conservation, the Hamiltonian can be written in the momentum-space as $\mathcal{H}_{\mathrm{sl-R}}  = \sum_{\vk}\Psi^{\dagger}_{\vk} \mathbf{H}_{\vk, \text{sl-R}}\Psi_{\vk}$ with 
\begin{align}\label{req.2}
\mathbf{H}_{\vk,\text{sl-R}} & =
\begin{pmatrix}
\mathbf{H}_{\vk,\text{sl}} (\lambda = 0) & \mathbf{H}_{\vk, \text{R}} \\
\mathbf{H}^{\dagger}_{\vk, \text{R}} & \mathbf{H}_{\vk,\text{sl}} (\lambda = 0)  \\
\end{pmatrix},
\end{align}
where $\mathbf{H}_{\vk, \text{R}}$ is a $3 \times 3$ Hamiltonian written as
\begin{equation}\label{req.3}
\mathbf{H}_{\vk, \text{R}} = \begin{pmatrix}
0 & -i\lambda_{\mathrm{R}}\sin \frac{k_1}{2} & -\lambda_{\mathrm{R}}\sin \frac{k_2}{2} \\
i\lambda_{\mathrm{R}}\sin \frac{k_1}{2} & 0 & 0 \\
-\lambda_{\mathrm{R}}\sin \frac{k_2}{2} & 0 & 0
\end{pmatrix}.
\end{equation}
The spectrum for the monolayer system
%===========================================================================================
\begin{figure}[b]
\centering
\includegraphics[width=1\linewidth]{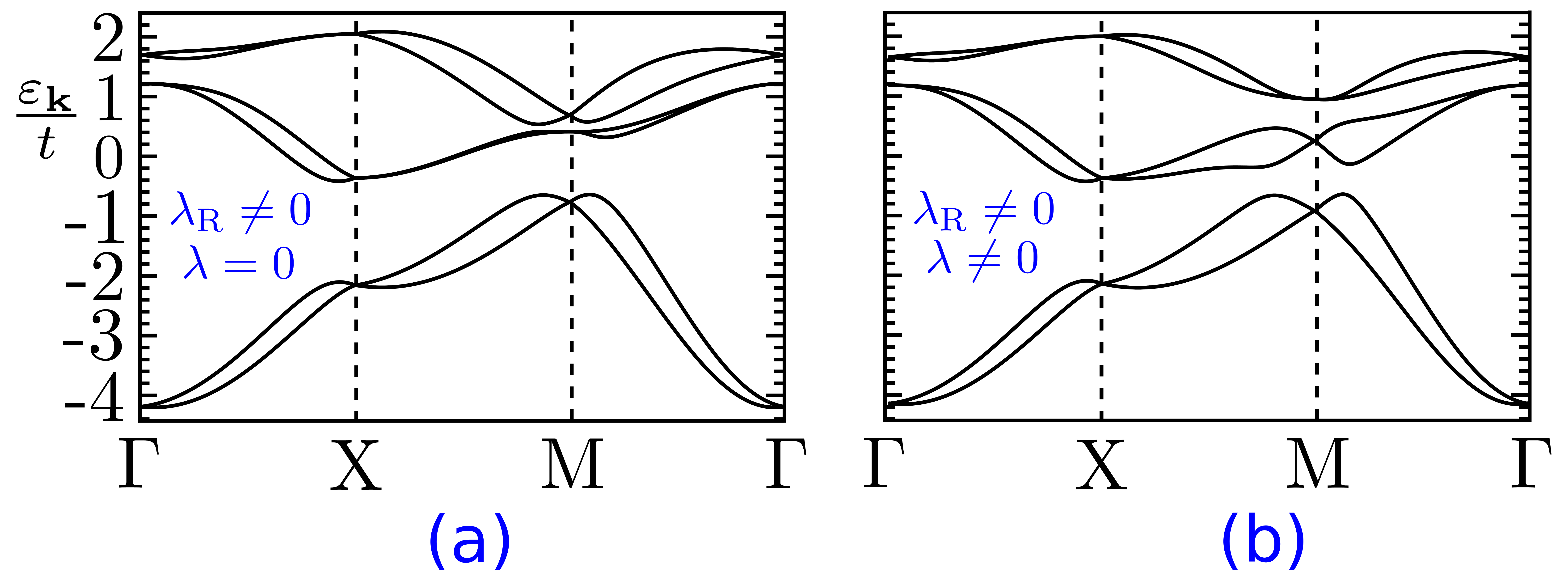} 
\caption{(a) The spectrum for the monolayer Lieb lattice with Rashba spin-orbit coupling. The RSOC strength is chosen to be $\lambda_{\mathrm{R}} = 0.4$ eV. There are six bands as the spin conservation is destroyed. (b) The spectrum in the presence of both the Rashba and intrinsic SOC ($\lambda = 0.2$ eV). The ISOC further splits the bands. The other TB parameters remain the same as in Fig.~\ref{fig:AFig1}.}
\label{fig:AFig6}
\end{figure} 
%===========================================================================================
with Rashba SOC is obtained by diagonalizing the Hamiltonian in Eq.~\ref{req.2}. The band structure is shown in Fig.~\ref{fig:AFig6}a. As RSOC breaks the spin degeneracy, there are six bands as compared to the three bands in Fig.~\ref{fig:AFig1}. For completeness, we also show the band structure with both the Rashba and intrinsic SOC. 

%=================================================================================================================================
\begin{figure}[t]
\centering
\includegraphics[width=1\linewidth]{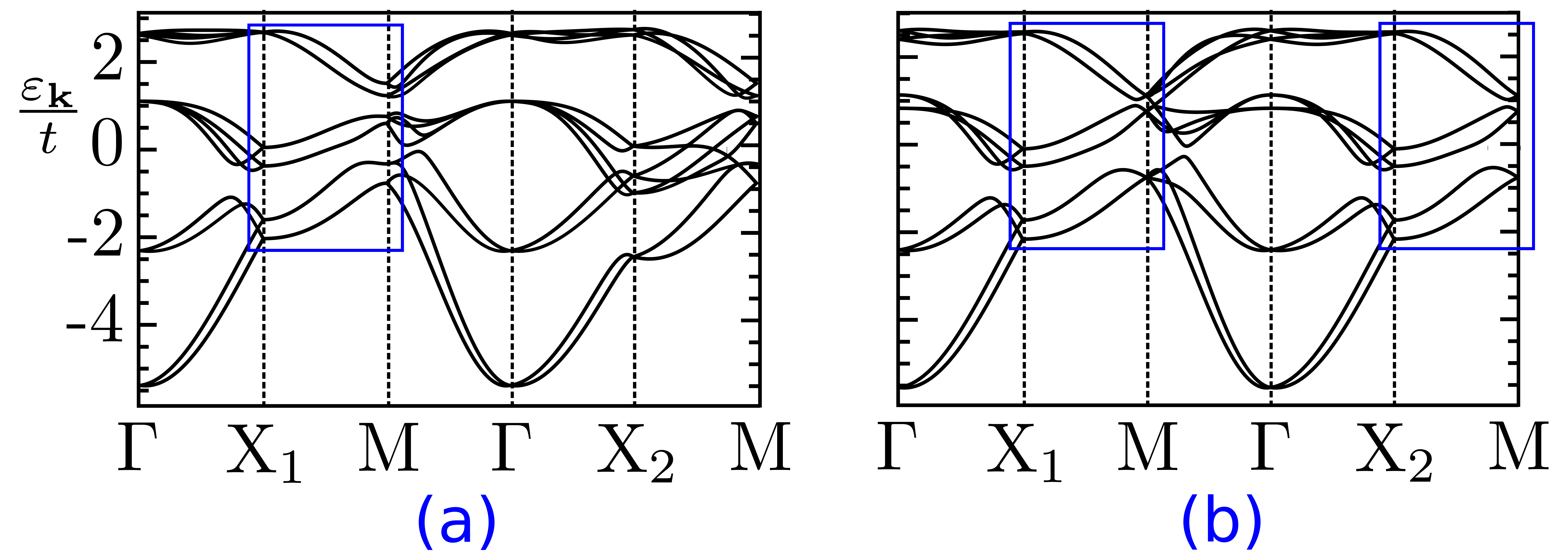} 
\caption{The band structure for the (a) AB- and (b) ABC-stacked bilayer Lieb lattice in the presence of both the Rashba and intrinsic SOC. The TB parameters for each layer are chosen the same as before in Fig.~\ref{fig:AFig6}, whereas the interlayer couplings remain the same as in Fig.~3 in the main text. The degenerate QBCL bands are present along the BZ edge.}
\label{fig:AFig7}
\end{figure} 
%=================================================================================================================================  

Finally, we provide the details of the analysis for the non-symmorphically stacked bilayer systems: AB- and ABC-stacked Lieb lattice with both the Rashba and intrinsic SOC. Surprisingly, the QBCL structure remains preserved even in the presence of Rashba SOC in both the layers. Note that, we consider the sign of the RSOC ($\lambda_{\mathrm{R}}$) to be the same in both the layers. The corresponding spectrum for AB- and ABC-stacked bilayer Lieb lattice is shown in Fig.~\ref{fig:AFig7}a,b. Again the broken spin-degeneracy due to the Rashba SOC leads to six (instead of three) QBCLs along the BZ edge.

%=========================================================================
\section{AA-stacked Bilayer Lieb Lattice \label{sec:app3}}
%=========================================================================

\begin{figure}[t]
\centering
\includegraphics[width=1\linewidth]{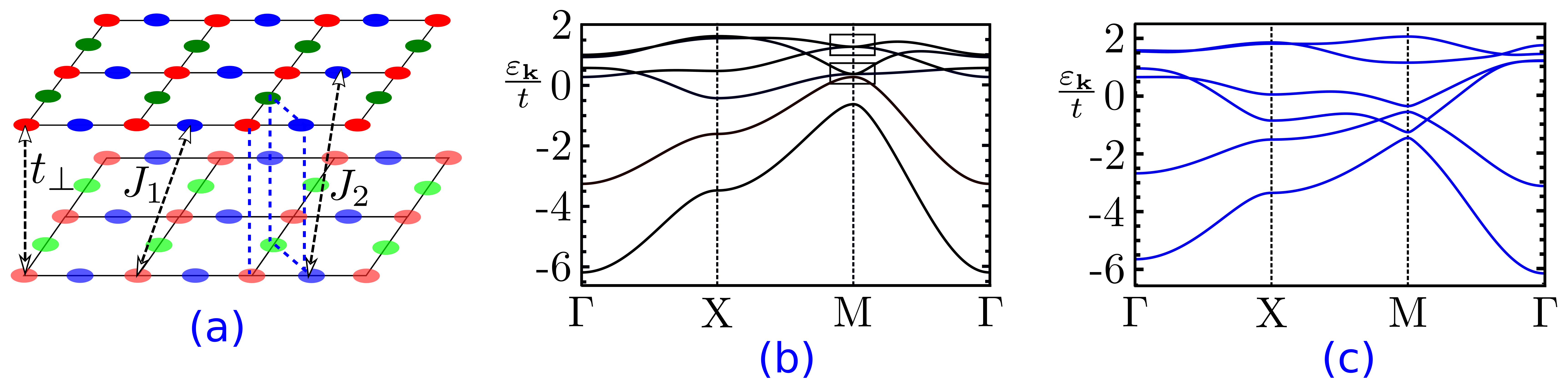} 
\caption{(a) An illustration of the bilayer Lieb lattice with conventional AA-stacking. The unit-cell for the effective 2D bilayer lattice is composed of three dimer sites as marked by the dashed lines. The inter-layer couplings ($t_{\perp},J_1,J_2$) are assumed between the dimer sites. The band structure [without (b) and with (c) the intrinsic spin-orbit coupling] along the high-symmetry directions in the BZ, obtained by diagonalizing the Hamiltonian in Eq.~\ref{eq.3a}. The tight-binding parameters for each layer are the same as in Fig.~\ref{fig:AFig1}e, with non-vanishing inter-layer couplings $t_\perp = 0.45$, $J_1=0.25$ and $J_2 = 0.15$. All values are in the units of eV. The spin-orbit coupling strength $\lambda = 0.35$ in panel (c).}
\label{fig:AFig2}
\end{figure}

In the main text, we focused on two unique stackings: AB and ABC with the emergent non-symmorphic structure. Here, we analyze the properties of the band-structure for the conventional AA-stacked bilayer Lieb lattice. In comparison to Eq. (2) in the main text, the corresponding Hamiltonian is written as
\begin{subequations}
\begin{align} 
& \mathbf{H}_{\vk,\text{aa}} = \mathbf{H}_{\vk,\text{sl}}\otimes \mathbb{I}_{\sigma} + \mathbf{V}_{\vk,\text{aa}}\otimes \sigma_1,  \label{eq.3a} \\
\mathbf{V}_{\vk,\text{aa}} & = -t_{\perp} c_1\mathbb{I}_3 - 2J_1c_1 \mathsf{\Gamma}_1 - 2J_1c_2 \mathsf{\Gamma}_4 - 4J_2c_1c_2 \mathsf{\Gamma}_6 , \label{eq.3b}\\
\mathsf{\Gamma}_1  &= 
  \begin{pmatrix} 
  0 & 1 & 0 \\
  1 & 0 & 0 \\
  0 & 0 & 0
  \end{pmatrix},~~
\mathsf{\Gamma}_4  =
 \begin{pmatrix}
  0 & 0 & 1 \\
  0 & 0 & 0 \\
  1 & 0 & 0
 \end{pmatrix}, ~~
\mathsf{\Gamma}_6  =
 \begin{pmatrix}
  0 & 0 & 0 \\
  0 & 0 & 1 \\
  0 & 1 & 0
 \end{pmatrix}, 
\end{align} 
\end{subequations}
where $c_i = \cos k_i/2$ and $\mathsf{\Gamma}_i$'s are the traceless Gell-Mann matrices as discussed in the main text. To illustrate the evolution of the single-layer QBCP, we assume all the TB parameters in Eq.~\ref{eq.3a} to be non-zero and further consider non-vanishing interlayer couplings $t_{\perp}, ~J_1, ~J_2$. The unit-cell for AA-stacking is composed of three dimer sites as illustrated in Fig.~\ref{fig:AFig2}a. The band structure (without intrinsic spin-orbit coupling) is obtained by diagonalizing the Hamiltonian $\mathbf{H}_{\vk,\text{aa}}$. The spectrum for AA-stacked bilayer Lieb lattice is shown in Fig.~\ref{fig:AFig2}b.  Quite intuitively, we notice that the individual QBCP in each layer eventually generates two distinct QBCPs at the M point of the BZ. Based on this result, we conclude that an AA-stacked $\mathcal{N}$-layer Lieb lattice hosts distinct $\mathcal{N}$-QBCPs at the M point in the corresponding spectrum. The QBCPs become gapped and are lost when the intrinsic spin-orbit coupling is turned on for each layer. Unlike the other stackings discussed in detail in the main text, we do not observe any extended degeneracy along the BZ edge. It becomes evident that the QBCLs are only generic features of the emergent non-symmorphic structure in the case of AB- and ABC-stackings.
%=================================================================================================================================
%=================================================================================================================================  

%====================================================================================================
\section{Symmetry analysis for the degenerate QBCLs in bilayer Lieb lattice: AB- \& ABC-stacking \label{sec:app4}} 
%====================================================================================================

%=================================================================================================================================
\begin{figure}[b]
\centering
\includegraphics[width=1\linewidth]{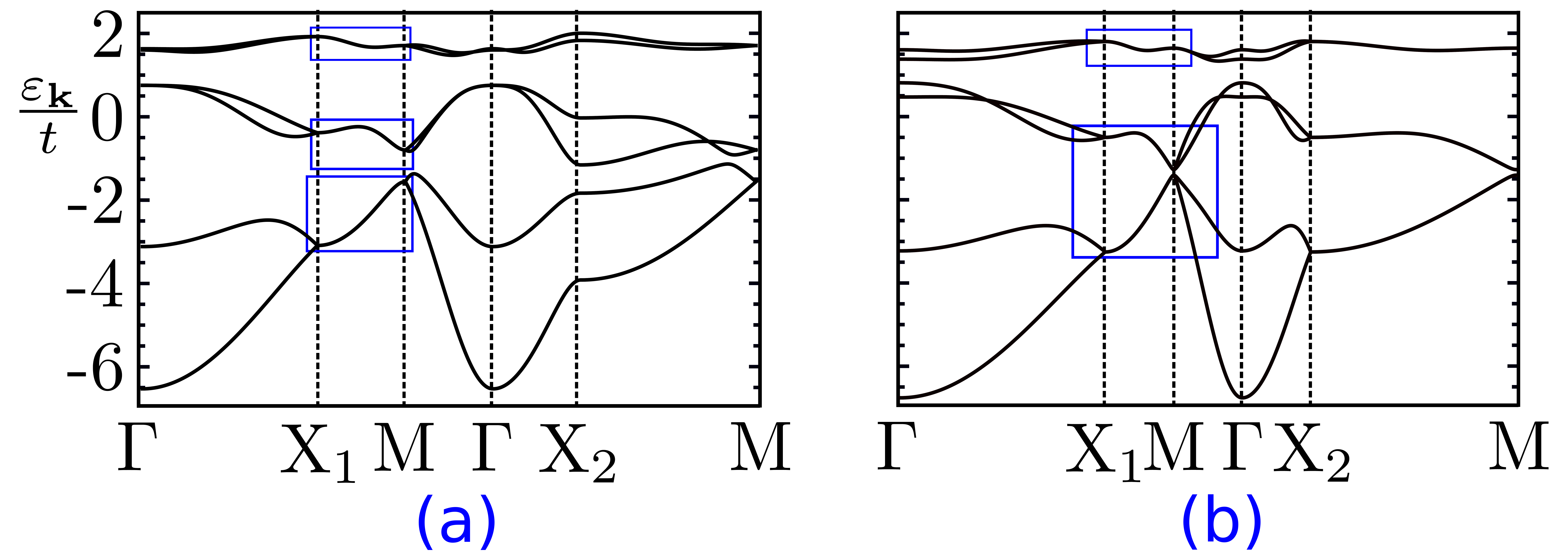} 
\caption{The spectrum for (a) AB- and (b) ABC-stacked bilayer Lieb lattice under uniaxial strain along the horizontal direction. In both the cases, the QBCL bands exist along the edge of the Brillouin zone. For AB-stacking the degeneracy extends only along X$_1$ $\rightarrow$ M (reduced path because of the uniaxial strain). The tight-binding parameters are chosen to be the same as in Fig.~3 in the main text.}
\label{fig:AFig5}
\end{figure} 
%================================================================================================================================
%=================================================================================================================================
\begin{figure*}[htb!]
\centering
\includegraphics[width=1\linewidth]{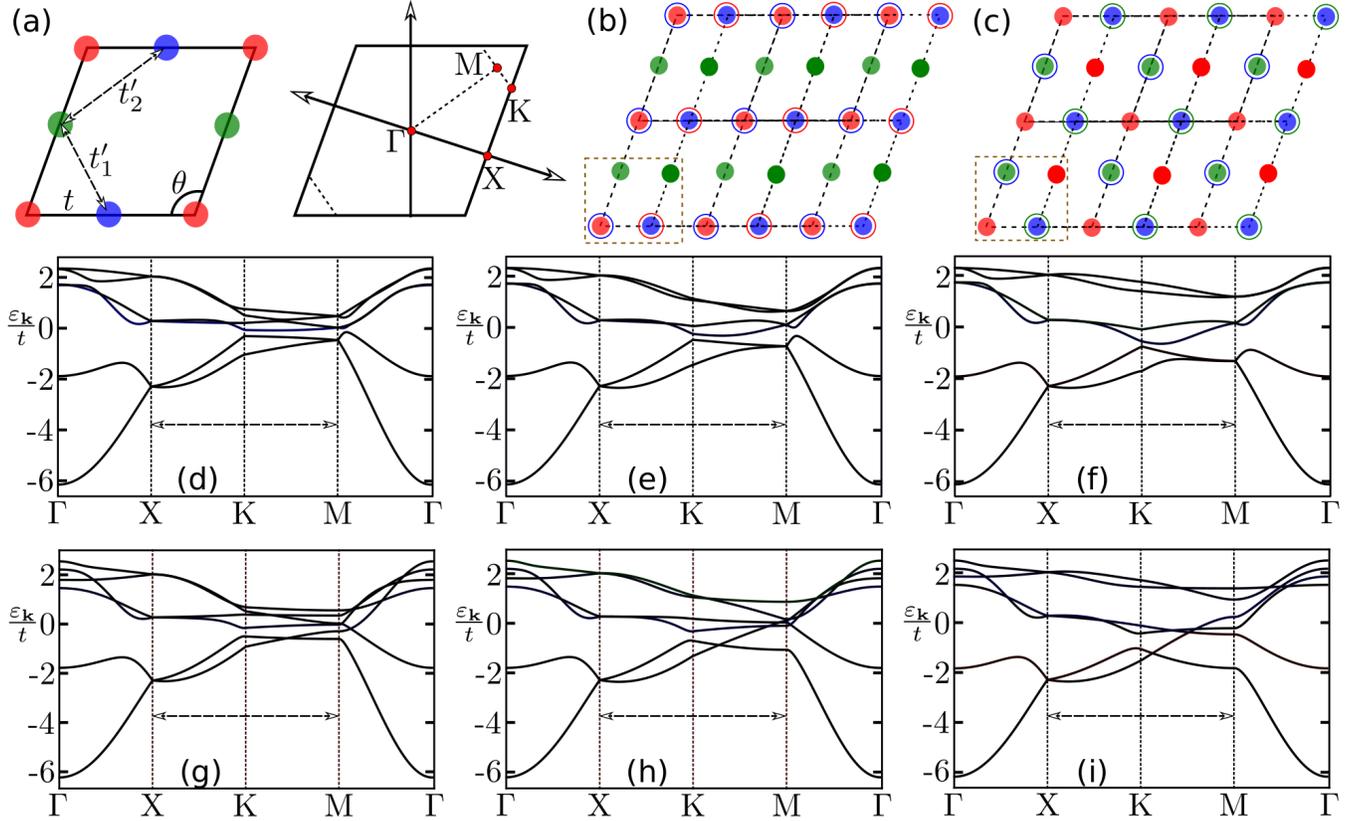} 
\caption{(a) A pictorial representation of a single-layer oblique Lieb lattice \cite{PhysRevB.99.125131} and its corresponding Brillouin zone. The high-symmetry points are denoted by the filled red circles in the BZ. (b-c) A schematic of the bilayer oblique Lieb lattice stacked in AB- and AB$\Theta$- configurations, respectively (c-axis top view). The monomer sites are labeled by filled circles whereas the filled double circles signify the dimer sites. The outer color marks which sublattice site is on the top layer. The color coding for each layer is chosen in the same way as in Fig.~\ref{fig:AFig2}a. (d-f) Evolution of the band structure as a function of the oblique angle $\theta$ ($97.5^{\circ} \rightarrow 105^{\circ} \rightarrow 112.5^{\circ}$) for the AB-stacking. (g-i) Similar evolution for the AB$\Theta$-stacking. The tight-binding parameters are assumed as $t = 1, t_{\perp} = 0.45, J_1 = 0.25$. The other parameters $t'_1,t'_2, J^B_2, J^C_2$ are interpolated between the Lieb and Kagome limits. The Brillouin zone edge is illustrated by the region between X and M points (dashed arrow). All values are in the units of eV.}
\label{fig:AFig3}
\end{figure*} 
%================================================================================================================================

In the main text, we mentioned that the degenerate QBCL structure is probably tied to the fractional glide transformations for AB- and ABC-stakced bilayer Lieb systems. In order to analyze the robustness of the QBCL bands, we employ various distortions to the non-symmorphically stacked bilayer Lieb systems. First, we apply a uni-axial strain to the lattice. In this case, the apparent $C_4$-rotation symmetry is reduced to a $C_2$-rotation symmetry. Surprisingly, in this case the spectrum still contains the degenerate QBCL bands along the edge of the BZ (see Fig.~\ref{fig:AFig5}).

However, for an oblique analogue of the stacked Lieb systems, the QBCLs are completely destroyed in the band structure. To demonsrate this, we consider two coupled oblique Lieb lattices in the two Bernal-stacked configurations (AB and AB$\Theta$) as shown in Fig.~\ref{fig:AFig3}b,c. The filled and double-filled circles in the unit-cell label the monomer and the dimer sites, respectively. For the double-filled circles, the outer color specifies atoms on the top layer. Each monolayer oblique Lieb structure, characterized by the angle $\theta$ as in Fig.~\ref{fig:AFig3}a, is obtained by applying a continuous shear along the $(11)$ direction to an ideal Lieb lattice. For $\theta = 90^{\circ}$ we obtain the Lieb lattice, while for $\theta = 120^{\circ}$ we generate the Kagome lattice. Here, we build upon the band structure calculation for the single layer oblique Lieb lattice in Ref.~\onlinecite{PhysRevB.99.125131}, and show that an arbitrary small shear destroys the degeneracy lines (see Fig.~\ref{fig:AFig3}d-i).
  
 For simplicity, the single-layer Hamiltonian for the oblique Lieb lattice is constructed in the presence of only two tight-binding parameters: nearest-neighbor hopping $t$ and the next-nearest neighbor hoppings $t'_1,t'_2$, respectively~\cite{PhysRevB.99.125131}. As $\theta$ increases from $90^{\circ}$ to $120^{\circ}$ the hopping $t'_1$ increases and $t'_2$ decreases. The lattice unit vectors are defined as $\bm a_1 = (1,0)$ and $\bm a_2 = (-\cos \theta, \sin \theta)$. The corresponding hexagonal parallelogon BZ is shown in Fig.~\ref{fig:AFig3}a. The reciprocal lattice vectors are obtained as $\bm b_1 = 2\pi(1,\cot \theta)$ and  $\bm b_2 = 2\pi(0, \csc \theta)$. Consequently, the $\Gamma \rightarrow$ X path is determined by $\pi (1,\cot\theta)$. The path from the X to the K point is obtained by finding the vector perpendicular to the previous vector as $(-\cos \theta, \sin \theta)$. However, pinpointing the K point in the oblique Lieb BZ is a little tricky. To determine the K point, we first find the M point which is easily obtained as
\begin{equation}\label{seq.3}
\text{M} = \frac{\vec{\bm b}_1 + \vec{\bm b}_2}{2} = \pi (1, \cot \theta + \csc \theta).
\end{equation} 
Now, we find a vector which is perpendicular to the one connecting the $\Gamma$ point to the M point. The goal is now to determine the intersecting point between this vector and the vector along the X to the K point. The latter one is easily obtained from the vector $\Gamma \rightarrow$ X. Consequently, we obtain the K point as
\begin{equation}\label{seq.4}
\text{K} = \pi \left( \frac{1-2\cos \theta}{1-\cos\theta},\cot \theta + \cot \frac{\theta}{2} \right).
\end{equation} 
The modified single layer Hamiltonian is written as
\begin{widetext}
\begin{equation}\label{seq.5}
\mathbf{H}_{\vk,\text{sl}} = \begin{pmatrix}
0 & -2t\cos \frac{k_1}{2} & -2t\cos \frac{k_2}{2} \\
-2t\cos \frac{k_1}{2} &  0 &  -2t'_1 \cos(\frac{k_1+k_2}{2}) -2t'_2 \cos(\frac{k_1-k_2}{2})\\
-2t\cos \frac{k_2}{2} & -2t'_1 \cos(\frac{k_1+k_2}{2}) -2t'_2 \cos(\frac{k_1-k_2}{2}) & 0 \\
\end{pmatrix},
\end{equation}
\end{widetext}
where $k_i = \vk \cdot \bm{a}_i$ and the parameters $t,t'_1,t'_2$ have been defined earlier. In a similar fashion to the intra-layer hoppings $t'_1,t'_2$, we assume different inter-layer couplings $t_{\perp}, J_1, J^B_2, J^C_2$ (not shown in Fig.~\ref{fig:AFig3}b,c). The first two couplings $t_{\perp},J_1$ are defined in the same way as in the main text, whereas the remote couplings $J^{B,C}_2$ become dependent on the angle $\theta$: $J^B_2$ monotonically increases to $J_1$ and $J^C_2$ keeps decreasing as $\theta$ varies between $90^{\circ}$ to $120^{\circ}$. The bilayer coupling Hamiltonian for AB- and AB$\Theta$-stackings are defined as 
\begin{widetext}
\begin{subequations}
\begin{align} 
& \mathbf{H}_{\vk,\text{ab/ab}\theta} = \mathbf{H}_{\vk,\text{sl}}\otimes \mathbb{I}_{\sigma} + \mathbf{V}_{\vk,\text{ab/ab}\theta}\otimes \sigma_1,  \label{eq.4a} \\
\mathbf{V}_{\vk,\text{ab}} & = \begin{pmatrix}
2J_1\cos \frac{k_1}{2} & t_{\perp} & S_\vk\\
t_{\perp} &  2J_1\cos \frac{k_1}{2}  2J_1\cos \frac{k_2}{2} \\
S_\vk & 2J_1\cos \frac{k_2}{2}  & 2J_1\cos \frac{k_1}{2}\\
\end{pmatrix},  \;   \mathbf{V}_{\vk,\text{ab}\theta}  = \begin{pmatrix}
S_\vk & 2J_1\cos \frac{k_2}{2} & 2J_1 \cos \frac{k_1}{2}\\
2J_1\cos \frac{k_2}{2} & S_\vk & t_{\perp} \\
2J_1\cos \frac{k_1}{2} & t_{\perp}  & S_\vk\\
\end{pmatrix},
\end{align} 
\end{subequations}
\end{widetext}
where $S_\vk$ is defined according to
\begin{equation}\label{appeq.5}
S_\vk = -2 J^B_2 \cos \left(\frac{k_1+k_2}{2}\right)-2 J^C_2 \cos \left(\frac{k_1-k_2}{2}\right).
\end{equation}
The band structure for the two different Bernal-type stackings are obtained by diagonalizing the bilayer Hamiltonian (Eq.~\ref{eq.4a}). The corresponding spectra for the two stackings are shown in the middle and bottom panels in Fig.~\ref{fig:AFig3}, respectively. The broken $C_4$ rotation symmetry in both cases results into gapped bands along the BZ edge: X $\rightarrow$ K $\rightarrow$ M, where K is the edge point in the oblique BZ (Fig.~\ref{fig:AFig3}a). The variations of the spectrum for different oblique angles $\theta = 97.5^{\circ},105.5^{\circ},112.5^{\circ}$ are shown in Fig.~\ref{fig:AFig3}d-f and Fig.~\ref{fig:AFig3}g-i for AB- and AB$\Theta$-stackings, respectively. For $\theta = 120^{\circ}$, we obtain a bilayer Kagome structure and reproduce the spectrum analyzed in Ref.~\onlinecite{PhysRevB.100.155421}. As $\theta$ decreases progressively, the gap between the pair of bands along the BZ edge also decreases and eventually vanishes at $90^{\circ}$ where the QBCLs {\it reappear} as in Fig.~3a,b, in the main text. 

Based on the above analysis, we anticipate that the degenerate QBCL bands are a generic feature of non-symmorphically stacked 2D Bravais lattices with dihedral point group symmetry~\cite{PhysRevB.101.165121}.
 
%============================================================
\section{Multilayer Lieb lattice: Wilson Loop analysis \label{sec:app5}}
%============================================================

\begin{figure}[t]
\centering
\includegraphics[width=1\linewidth]{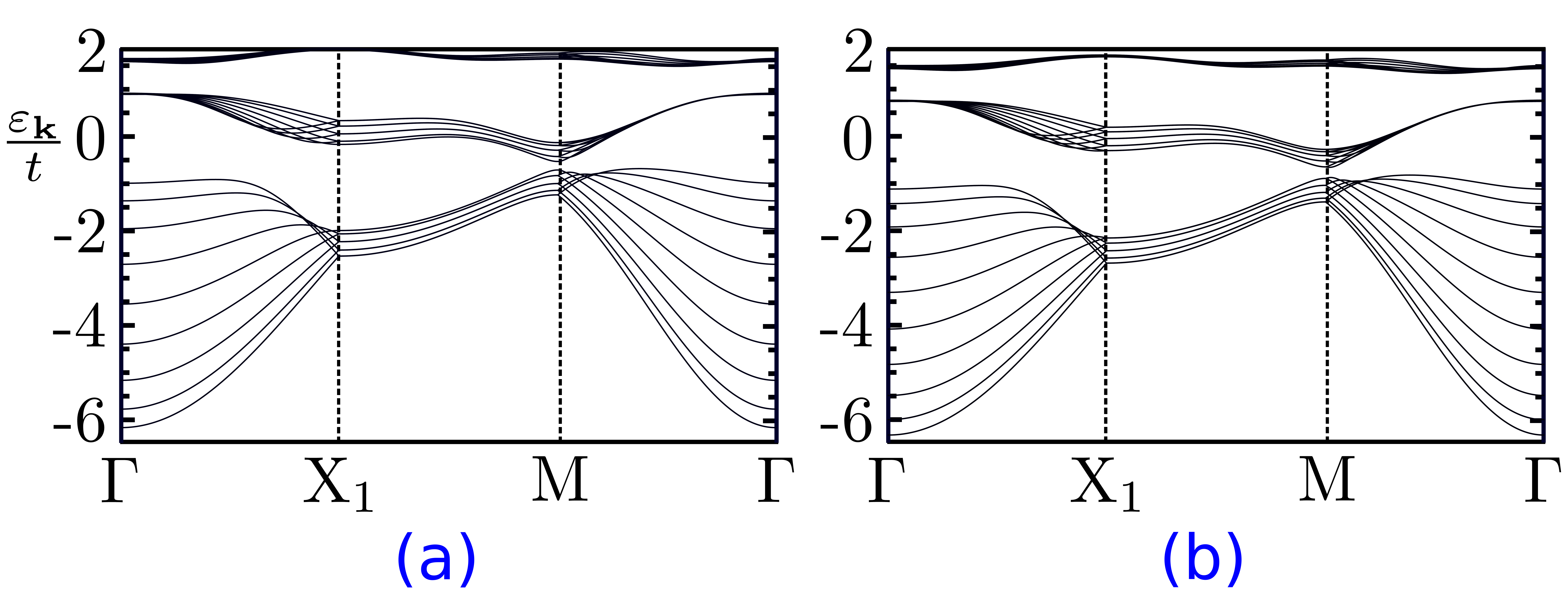} 
\caption{The band structure for the (a) nine- and (b) ten-layer stacked Lieb lattice system in the presence of only intra-layer intrinsic-spin orbit coupling $\lambda$. The stacking is considered of only AB-type between any two adjacent layers. The tight-binding parameters used are chosen to be the same as used in the main text. (a) There are four QBCLs and one lone band in each band-subspace, while (b) each band-subspace contains only QBCLs (five in this case) and no lone band.}
\label{fig:AFig4}
\end{figure}

In this section, we provide the tight-binding analysis for the multilayer stacked Lieb lattices. As the properties of the band structure for the bilayer stacked lattice are simply inherited in the multilayer structure, we primarily focus on the band structure for the AB-stacking. Here, we show the dispersion for nine and ten layer stackings.  For each case, the stacking is considered to be of the AB-type in between any two adjacent layers [we consider the interlayer coupling again the same as ($t_{\perp},J_1,J_2$)]. The band structures for the two cases are shown in Fig.~\ref{fig:AFig4}a,b for 9 and 10 layers, respectively. Again, we observe that within a finite region in the parameter space, the spectrum consists of three gapped band-subspaces and each of the subspaces contains QBCLs. For odd number of layers (nine-layer or nonalayer stacked case) there are four QBCLs and one lone band, whereas for even number of layers (ten-layer or decalayer stacked case), the subspace only consists of QBCLs (in this case the number is five). Of course, the same qualitative features hold true for the corresponding ABC-stacked cases (not shown here). We compute the Wilson loop spectrum for each of the band-subspaces and analyze the winding of non-abelian Berry phases (in the same way as explained in the main text). Consequently, we obtain the Chern number distribution for these two cases arranged as $\mathcal{C} = (0,9,-9)$ and $\mathcal{C} = (0,10,-10)$, for nine- and ten-layers, respectively.

%=================================================================================================================================
%================================================================================================================================

We finally illustrate the key steps of the computation of the Wilson loops. As stressed in the main text, the gauge fixing is only needed for the two end points of a Wilson loop. Consequently, we consider a periodic gauge and incorporate such features by suitably constructing a matrix $\mathcal{T_{\mathbf{G}}}$~\cite{PhysRevB.95.241101} such that the following constraint is satisfied
\begin{equation}\label{eq.transfer_matrix}
\mathcal{T_{\mathbf{G}}}\cdot \mathcal{H}_{\vk+\mathbf{G}} = \mathcal{H}_{\vk} \cdot \mathcal{T_{\mathbf{G}}},
\end{equation}
where $\mathbf{G}$ is the reciprocal lattice vector and $\mathcal{H}_\vk$ is the corresponding Hamiltonian of the underlying system. In the last step of the Wilson loop computation, we consider the following Bloch-function at the end point of the loop as
\begin{equation}\label{eq.final}
\ket{u_m(\vk + \mathbf{G})} = \mathcal{T_{\mathbf{G}}} \cdot \ket{u_m(\vk)}. 
\end{equation} 
%%%%%%%%%%%%%%%%%%%%%%%%%%%%%%%%%%%%%%%%%%%%%%%%%%%%%%%%%%%%%%%%%%%%%%%%%%%%%%%%%%%%%%%%%%%%%%%%%%%%%%%%%%%%%%%%%%%

%%%%%%%%%%%%%%%%%%%%%%%%%%%%%%%%%%%%%%%%%%%%%%%%%%%%%%%%%%%%%%%%%%%%%%%%%%%%%%%%%%%%%%%%%%%%%%%%%%%%%%%%%%%%%%%%%%%%%%%%%%%%%%%%
\bibliography{References}
\bibliographystyle{apsrev4-1}
%%%%%%%%%%%%%%%%%%%%%%%%%%%%%%%%%%%%%%%%%%%%%%%%%%%%%%%%%%%%%%%%%%%%%%%%%%%%%%%%%%%%%%%%%%%%%%%%%%%%%%%%%%%%%%%%%%%%%%%%%%%%%%%%
\end{document}